\documentclass[]{aastex631}

\newcommand{\HI}{H{\sc i}}
\newcommand{\mhi}{M$_{\rm HI}$}

\usepackage{graphicx}
\usepackage{amsmath}

\newcommand{\myedit}{}
\shorttitle{WALLABY: HI in the host galaxy of FRB20211127}
\shortauthors{Glowacki et al.}
\graphicspath{{./}{figures/}}

\begin{document}

\title{WALLABY Pilot Survey: HI in the host galaxy of a Fast Radio Burst}

\correspondingauthor{Marcin Glowacki}
\email{marcin.glowacki@curtin.edu.au}

\author[0000-0002-5067-8894]{M. Glowacki}
\affiliation{International Centre for Radio Astronomy Research (ICRAR), Curtin University, Bentley, WA 6102, Australia}

\author[0000-0003-4844-8659]{K. Lee-Waddell}
\affiliation{International Centre for Radio Astronomy Research (ICRAR), The University of Western Australia, 35 Stirling Hwy, Crawley, WA 6009, Australia}
\affiliation{CSIRO Space and Astronomy, PO Box 1130, Bentley, WA 6102, Australia}
\affiliation{International Centre for Radio Astronomy Research (ICRAR), Curtin University, Bentley, WA 6102, Australia}

\author{A.~T.~Deller}
\affiliation{Centre for Astrophysics and Supercomputing, Swinburne University of Technology, Hawthorn, VIC, 3122, Australia}

\author{N. Deg}
\affiliation{Department of Physics, Engineering Physics, and Astronomy, Queen's University, Kingston, ON, K7L 3N6, Canada}

\author{A. C. Gordon}
\affiliation{Center for Interdisciplinary Exploration and Research in Astrophysics and Department of Physics and Astronomy,
Northwestern University, 2145 Sheridan Road, Evanston, IL 60208-3112, USA}

\author{J. A. Grundy}
\affiliation{International Centre for Radio Astronomy Research (ICRAR), Curtin University, Bentley, WA 6102, Australia}
\affiliation{CSIRO Space and Astronomy, PO Box 1130, Bentley, WA 6102, Australia}

\author{L. Marnoch}
\affiliation{School of Mathematical and Physical Sciences, Macquarie University, NSW 2109, Australia}
\affiliation{Astronomy, Astrophysics and Astrophotonics Research Centre, Macquarie University, Sydney, NSW 2109, Australia}
\affiliation{CSIRO, Space and Astronomy, PO Box 76, Epping NSW 1710 Australia}
\affiliation{ARC Centre of Excellence for All Sky Astrophysics in 3 Dimensions (ASTRO 3D), Australia}

\author{A. X. Shen}
\affiliation{CSIRO Space and Astronomy, PO Box 1130, Bentley, WA 6102, Australia}

\author{S. D. Ryder}
\affiliation{School of Mathematical and Physical Sciences, Macquarie University, NSW 2109, Australia}
\affiliation{Astronomy, Astrophysics and Astrophotonics Research Centre, Macquarie University, Sydney, NSW 2109, Australia}

\author{R.~M.~Shannon}
\affiliation{Centre for Astrophysics and Supercomputing, Swinburne University of Technology, Hawthorn, VIC, 3122, Australia}

\author{O. I. Wong}
\affiliation{CSIRO Space and Astronomy, PO Box 1130, Bentley, WA 6102, Australia}
\affiliation{International Centre for Radio Astronomy Research (ICRAR), The University of Western Australia, 35 Stirling Hwy, Crawley, WA 6009, Australia}
\affiliation{ARC Centre of Excellence for All Sky Astrophysics in 3 Dimensions (ASTRO 3D), Australia}

\author[0000-0002-9214-8613]{H. Dénes}
\affiliation{ASTRON, Netherlands Institute for Radio Astronomy, Oude Hoogeveensedĳk 4, 7991 PD Dwingeloo, The Netherlands}

\author[0000-0003-4351-993X]{B. S. Koribalski}
\affiliation{CSIRO Astronomy and Space Science, Australia Telescope National Facility, P.O. Box 76, NSW 1710, Australia}
\affiliation{School of Science, Western Sydney University, Locked Bag 1797, Penrith, NSW 2751, Australia}

\author{C. Murugeshan}
\affiliation{CSIRO Space and Astronomy, PO Box 1130, Bentley, WA 6102, Australia}
\affiliation{ARC Centre of Excellence for All Sky Astrophysics in 3 Dimensions (ASTRO 3D), Australia}

\author{J. Rhee}
\affiliation{International Centre for Radio Astronomy Research (ICRAR), The University of Western Australia, 35 Stirling Hwy, Crawley, WA 6009, Australia}
\affiliation{ARC Centre of Excellence for All Sky Astrophysics in 3 Dimensions (ASTRO 3D), Australia}

\author{T. Westmeier}
\affiliation{International Centre for Radio Astronomy Research (ICRAR), The University of Western Australia, 35 Stirling Hwy, Crawley, WA 6009, Australia}
\affiliation{ARC Centre of Excellence for All Sky Astrophysics in 3 Dimensions (ASTRO 3D), Australia}

\author{S. Bhandari}
\affiliation{ASTRON, Netherlands Institute for Radio Astronomy, Oude Hoogeveensedĳk 4, 7991 PD Dwingeloo, The Netherlands}
\affiliation{Joint institute for VLBI ERIC, Oude Hoogeveensedĳk 4, 7991 PD Dwingeloo, The Netherlands}
\affiliation{Anton Pannekoek Institute for Astronomy, University of Amsterdam, Science Park 904, 1098 XH, Amsterdam, The Netherlands}
\affiliation{CSIRO, Space and Astronomy, PO Box 76, Epping NSW 1710 Australia}

\author{A. Bosma}
\affiliation{Aix Marseille Univ, CNRS, CNES, LAM, Marseille, France}

\author{B. W. Holwerda}
\affiliation{University of Louisville, Department of Physics and Astronomy, Natural Science Building 102, 40292 KY Louisville, USA}

\author{J. X. Prochaska}
\affiliation{Department of Astronomy and Astrophysics, University of California, Santa Cruz, CA 95064, USA}
\affiliation{Kavli Institute for the Physics and Mathematics of the Universe, 5-1-5 Kashiwanoha, Kashiwa 277-8583, Japan}
\affiliation{Simons Pivot Fellow}



\begin{abstract}

We report on the commensal ASKAP detection of a fast radio burst (FRB), FRB\,20211127{\myedit I}, and the detection of neutral hydrogen (\HI) emission in the FRB host galaxy, WALLABY\,J131913--185018 (hereafter W13--18). This collaboration between the CRAFT and WALLABY survey teams marks the fifth, and most distant, FRB host galaxy detected in \HI, not including the Milky Way. We find that W13--18 has a \HI\ mass of \mhi~=~6.5~$\times$~10$^{9}$ M$_{\odot}$, a \HI-to-stellar mass ratio of 2.17, and
coincides with a continuum radio source of flux density at 1.4~GHz of 1.3~mJy. The \HI\ global spectrum of W13--18 appears to be {\myedit asymmetric, albeit the \HI\ observation has a low S/N,} and the galaxy itself appears modestly undisturbed.
These properties are {\myedit compared to the} early literature of \HI\ emission detected in other FRB hosts to date, where either the \HI\ global spectra were strongly asymmetric, or there were clearly disrupted \HI\ intensity map distributions. W13--18 {\myedit lacks sufficient S/N to determine whether it is significantly less asymmetric in its \HI\ distribution than previous examples of FRB host galaxies. However, there are no strong signs of a major interaction in the 
optical image of the host galaxy}
that would stimulate a burst of star formation and hence the production of putative FRB progenitors related to massive stars and their compact remnants.

\end{abstract}


\section{Introduction} \label{sec:intro}

To date, we do not know the origin of fast radio bursts \cite[FRBs;][]{Lorimer2007}, the highly energetic radio pulses occurring on timescales of milliseconds and found to originate at extragalactic distances. While several theories exist \cite[see review by][]{Cordes2019}, this is an ongoing area of debate. To best address this issue, we need to better identify and dissect the host galaxies of FRBs, through localisation of their radio signal. One of the main aims of the Commensal Real-time ASKAP Fast Transients survey \cite[CRAFT;][]{Macquart2010,Bannister2017} is to localise FRBs {\myedit at} sub-arcsecond scales with the Australian Square Kilometre Array Pathfinder telescope \cite[ASKAP;][]{Deboer2009}. Dedicated follow-up observations (e.g. in optical wavelengths) of these localised FRB positions give us important information to categorise the host galaxies and better understand the possible mechanisms behind FRBs \citep{Bhandari2022}. 

The stellar information alone does not inform us on the \emph{gas} content and distribution, from which stars form and can lead to FRB progenitors. One way to map the gas distribution is through the neutral hydrogen (\HI) 21-cm transition with radio telescopes. The \HI\ spatial distribution and kinematics can constrain the recent history of a galaxy. Since \HI\ gas {\myedit often} extends {\myedit beyond} the stellar distribution, we can also better see the indicators of recent galaxy interactions, such as tidal tails and extensions \citep{Holwerda2011c,Reynolds2019}. Such features may be difficult to detect in optical studies, but can be prominent in \HI\ intensity maps and velocity fields. Another possible indicator of galaxy interactions can be seen in asymmetries of the \HI\ global spectrum \cite[e.g.][]{Deg2020}. 

\begin{table*}[t]
\small
\caption{Summary of other \HI\ emission detections for FRB host galaxies. Each host galaxy was concluded to have had a recent (or ongoing) galaxy merger or interaction event.}
\centering
\begin{tabular}{@{}llll@{}}
\hline\hline
FRB Name & Host Galaxy & Redshift & Notes on \HI\ content and host galaxy \\
 \hline
FRB\,20171020A & ESO\,601--G036 & 0.008672 & Lee-Waddell et al. (submitted). The host galaxy has a faint stellar \\
& & & companion and tidal tail visible in the 
\HI\ intensity map and a lopsided \HI\  \\ 
& & & global spectrum. Both features are attributed to a galaxy interaction event.\\
FRB\,20180916B & SDSS\,J0158+6542 & 0.03399 & \cite{Kaur2022} presented disturbed \HI\ distribution from intensity and \\
& & & velocity maps, and concluded this galaxy is merging.\\
FRB\,20181030A & NGC\,3252 & 0.0038 & \cite{Michalowski2021} found \HI\ global spectrum from \citet{Masters2014} to \\
& & & be strongly asymmetric relative to general galaxy population and GRBs. \\
FRB\,20200120E & M81 & 0.00014 & {\myedit M81 is the dominant galaxy of the M81/M82/NGC\,3077 group. The data} \\
& & & {\myedit in \cite{Chynoweth2008} shows a complex structure of the overall \HI\ }\\
& & & {\myedit distribution in this group, and the global spectrum of M81 by itself is }\\
& & & strongly asymmetric relative to general galaxy population and GRB hosts. \\
FRB\,20200428 & Milky Way & -- & Asymmetric \HI\ distribution covered in review by \cite{Kalberla2009}. \\
& & & Interacting with the Large Magellanic Cloud and Small Magellanic Cloud.\\
\hline\hline
\end{tabular}
\label{tab:comparison}
\end{table*}

Thus far, many identified FRB host galaxies show evidence of star formation and would hence be expected to contain \HI. However \HI\ emission has only been detected in four FRB host galaxies, outside of the Milky Way, due to the current limited number ($<30$) of localised FRBs and the large distances to their hosts limiting the ability to directly detect \HI\ emission (the median redshift of localised FRB hosts used in \citealp{James2022} is $z~=~0.237$). In Table~\ref{tab:comparison}, we summarise these galaxies with \HI\ emission detections. With the issues arising from low number statistics in mind, one early trend that has emerged is that all of these host galaxies were claimed to have strongly asymmetric 21-cm spectra or highly disturbed \HI\ distributions. \cite{Michalowski2021} highlights that the two published FRB host galaxies at the time that were known to {\myedit have \HI\ emission} -- not including the Galactic magnetar in the Milky Way -- had highly asymmetric \HI\ global spectra, to a far greater degree than what is typically seen for the general galaxy population or hosts of long-duration gamma-ray bursts (GRBs). While the \HI\ global spectrum analysed by \cite{Michalowski2021} for one FRB host is of the M81 group, fig.~2 of \cite{Chynoweth2008} showcases the strongly asymmetric spectrum for M81 in isolation. We also {\myedit note} that the \cite{Masters2014} spectrum of NGC~3252 analysed by \cite{Michalowski2021} includes a baseline ripple. Our Milky Way is 
currently interacting with other systems e.g. the Large and Small Magellanic Clouds. Recently, \cite{Kaur2022} reported a Giant Metrewave Radio Telescope (GMRT) detection of \HI\ in the host galaxy of the repeating FRB\,20180916B. They found that the \HI\ distribution was highly disturbed, the source had two tidal tails, and there was a clear \HI\ deficit between the centre of the galaxy and FRB location. Additionally, Lee-Waddell et al., {\myedit (submitted), using data from the Australia Telescope Compact Array, showed that} the identified host galaxy of FRB\,20171020A \citep{Mahony2018} has an asymmetric \HI\ global spectrum and a clear \HI\ tail. \cite{Hsu2023} recently presented asymmetric profiles of molecular gas (CO) in the host galaxy of FRB\,20180924B, and proposed FRBs could commonly appear in {\myedit kinematically disturbed} environments.

In all of these studies, the cause of the \HI\ global spectrum asymmetry or disturbed \HI\ distribution has been attributed to ongoing or recent galaxy interaction events. The tidal activity enhances star formation in the host, which supports ``fast FRB channels" \citep{Michalowski2021}, i.e. a massive star collapsing in a supernova explosion into a magnetar with a short delay time acting as the FRB progenitor \cite[as theorised for the FRB studied in][]{Kaur2022}. However, the current sample size is small and the statistics must be significantly {\myedit improved} to better understand the FRB progenitor environment. Is it possible that a massive star as the FRB progenitor could be created without the need for a galaxy interaction, e.g. simply within a spiral galaxy arm? Will this trend persist for \emph{all} FRB host galaxies observed in \HI\ or is the story more complicated?

In this paper, we report on the \HI\ content of one of these FRB hosts, FRB\,20211127{\myedit I}, {\myedit as derived from} commensal ASKAP observations of the Widefield ASKAP L-band Legacy All-sky Blind surveY \cite[WALLABY;][]{Koribalski2020} Pilot Survey Phase 2. During these observations, FRB\,20211127{\myedit I} was detected by the CRAFT survey and subsequently localised to a nearby galaxy. In Section~\ref{sec:observations}, we describe the WALLABY observations and source finding process. Section~\ref{sec:results} presents the analysis of the \HI\ properties. We compare with the existing literature of \HI\ in FRB host galaxies in Section~\ref{sec:comparison} and summarise our conclusions in Section~\ref{sec:conclusions}. 

\section{Data}\label{sec:observations}

\subsection{ASKAP observations}\label{sec:ASKAP}

The original WALLABY observations of regions in the vicinity of the NGC\,5044 galaxy group were taken in late November~2021 and spanned a frequency range of 1151.5 - 1439.5 MHz.
These observations had been affected by technical issues in the calibration data and could not be used for spectral line imaging. However, as part of the commensal observing strategy of CRAFT, an FRB was successfully detected with a signal-to-noise ratio S/N of 38 on 27 November 2021 UT 00:00:10, and localised to J2000 13:19:14.08, --18:50:16.7, with an estimated uncertainty of $0.2"$ in RA and $0.8"$ in Dec. The localisation made use of the astrometric pipeline initially described in \cite{Bannister2019} and extended in \cite{Day2020}{\myedit ; see also \cite{Scott2023}}. The S/N of the FRB in the post-processed image was 73, meaning that the positional uncertainty is dominated by the accuracy with which systematic offsets can be estimated and removed using background radio sources detected in an image of the field made using the 3.1s of voltage data (Deller et al., {in prep}). This source was found to be coincident with the galaxy WISEA\,J131913.96--185016.2 (also known as 
6dFGS~gJ131914.0-185017, WALLABY\,J131913--185018, and here after W13--18), with a redshift $z = 0.0469$ \citep{Jones2009} that is consistent with the dispersion measure DM~=~227~pc~cm$^{-3}$ via the Macquart Relation \citep{Macquart2020}. Additional optical follow-up of the FRB site and its host galaxy using $g$ and $I$ band imaging was performed with the FORS2 instrument mounted on Unit Telescope 1 (UT1) of the European Southern Observatory's Very Large Telescope (VLT). These data were reduced using standard pipelines to produce the images shown in Fig~\ref{fig:emissionoverlay}; further analysis will be forthcoming in Deller et al. (in prep).

After the calibration issues on ASKAP were rectified, the same regions, which included W13--18, were re-observed in early December 2021, under scheduling block (SB) IDs 34167 and 34278, as part of Pilot Phase 2 WALLABY observations. These data were edited, calibrated, and imaged using the automated ASKAPsoft processing pipeline \cite[version 1.6.2;][]{Whiting2020} and standard WALLABY processing parameters. Observations and data reduction were as described for the Phase 1 Pilot Survey \citep{Westmeier2022}. In short, the observations included 15,552 spectral channels across the frequency range of 1151.5--1439.5~MHz, with only the upper 144~MHz retained due to radio frequency interference (RFI) below 1300~MHz. Continuum emission was subtracted from the visibility data through a sky model derived from the calibrated and deconvolved continuum image, and imaging performed with a robust weighting of 0.5. Further continuum subtraction was then performed in the image domain. The multi-scale CLEAN algorithm was used for deconvolution. 
The one difference for Phase 2 data processing is using holography constrained beams for the primary beam correction, rather than circular Gaussian beams, for more accurate image fluxes. 

Each night of observations comprised one 36-beam footprint that was processed individually. Resulting data products are publicly available on CSIRO ASKAP Science Data Archive \cite[CASDA;][]{Huynh2020} under the associated SBIDs. The image products for corresponding footprints (i.e. footprints A and B with a common tile name) were mosaicked together to produce the final full sensitivity images and cubes. 
The combined WALLABY spectral line cube -- imaged using ASKAP baselines up to 2~km -- has a 30 arcsec synthesized beam, $\sim$4 km~s$^{-1}$ spectral resolution, and an RMS of $\sim$1.6 mJy beam$^{-1}$ per channel in the region around W13--18. We note this is the idealised RMS value for central beams in ASKAP’s FOV. The combined continuum map -- imaged using all ASKAP baselines -- has a 9 arcsec resolution and RMS of 20-30 $\mu$Jy beam$^{-1}$.

Source finding was carried out using the Source Finding Application 2 \cite[SoFiA 2;][]{Serra2015,Westmeier2021} through a custom pipeline developed by the Australian SKA Regional Centre (AusSRC), with a minimum S/N threshold of 3. \cite{Westmeier2022} provides a detailed description of the process used for WALLABY. W13--18 was identified, from the pipelined source finding, with a central position offset by 0.32$"$ and 2.8$"$ from the optical position of WISEA\,J131913.96--185016.2 \citep{Jarrett2000} in right ascension and declination respectively, consistent with the expected WALLABY centroid errors.

\section{Results and Discussion}

\subsection{HI and stellar properties}\label{sec:results}

\begin{table}[t]
\small
\caption{Measured properties of W13--18 in order of \HI\ flux, \HI\ mass, W50 (with error measured from generating synthetic spectra by perturbing each channel by the noise; see Section 3.2), stellar mass, radio continuum flux, total global SFR from the radio continuum {\myedit (total and after subtracting unresolved core emission)}, SFR from WISE mid-infrared photometry, the SFR of the last 30~Myr measured from the Prospector SED model, {\myedit the SFR from GALEX NUV photometry,} 1D asymmetry measurements (lopsidedness and residual of the integrated spectrum), and the 2D asymmetry measure.}
\centering
\begin{tabular}{ll}
\hline\hline
Quantity & Value \\
 \hline
$S_{\rm HI}$ & 0.63\,$\pm$0.1~Jy~km\,s$^{-1}$\\
$M_{\rm HI}$ & 6.5\,$\pm$ 1.3~$\times$~10$^{9}$~M$_{\odot}$\\
$W50$ & 150\,$\pm$ 19~km\,s$^{-1}$\\
$M_{*}$ & 3.0\,$\pm$0.1~$\times$~10$^{9}$~M$_{\odot}$ \\
$S_{\rm 1.37}$ & 1.3~mJy\\
SFR$_{\rm radio}$ & 2.91$\substack{+1.70 \\ -1.07}$\,$\substack{+0.05 \\ -0.05}$~$M_{\odot}$~yr$^{-1}$\\
SFR$_{\rm radio~no~core}$ & 1.77$\substack{+1.04 \\ -0.66}$\,$\substack{+0.07 \\ -0.07}$~$M_{\odot}$~yr$^{-1}$\\
SFR$_{\rm WISE}$ & 2.57\,$\pm$\,0.28 $M_{\odot}$\,yr$^{-1}$\\
SFR$_{\rm SED}$ & 0.45$\substack{+0.59 \\ -0.36}$\,M$_{\odot}$\,yr$^{-1}$\\
SFR$_{\rm UV}$ & 1.22\,$\pm$\,0.05 $M_{\odot}$\,yr$^{-1}$\\
$A_{\rm flux}$ & 1.12\,$\pm$\,0.13\\
$A_{\rm spec}$ & 0.296\,$\pm$\,0.091\\
$A_{2D}$ & 0.23\\
\hline
\hline
\end{tabular}
\label{tab:galdetails}
\end{table}

\begin{figure*}
\centering
\includegraphics[width=0.99\textwidth]{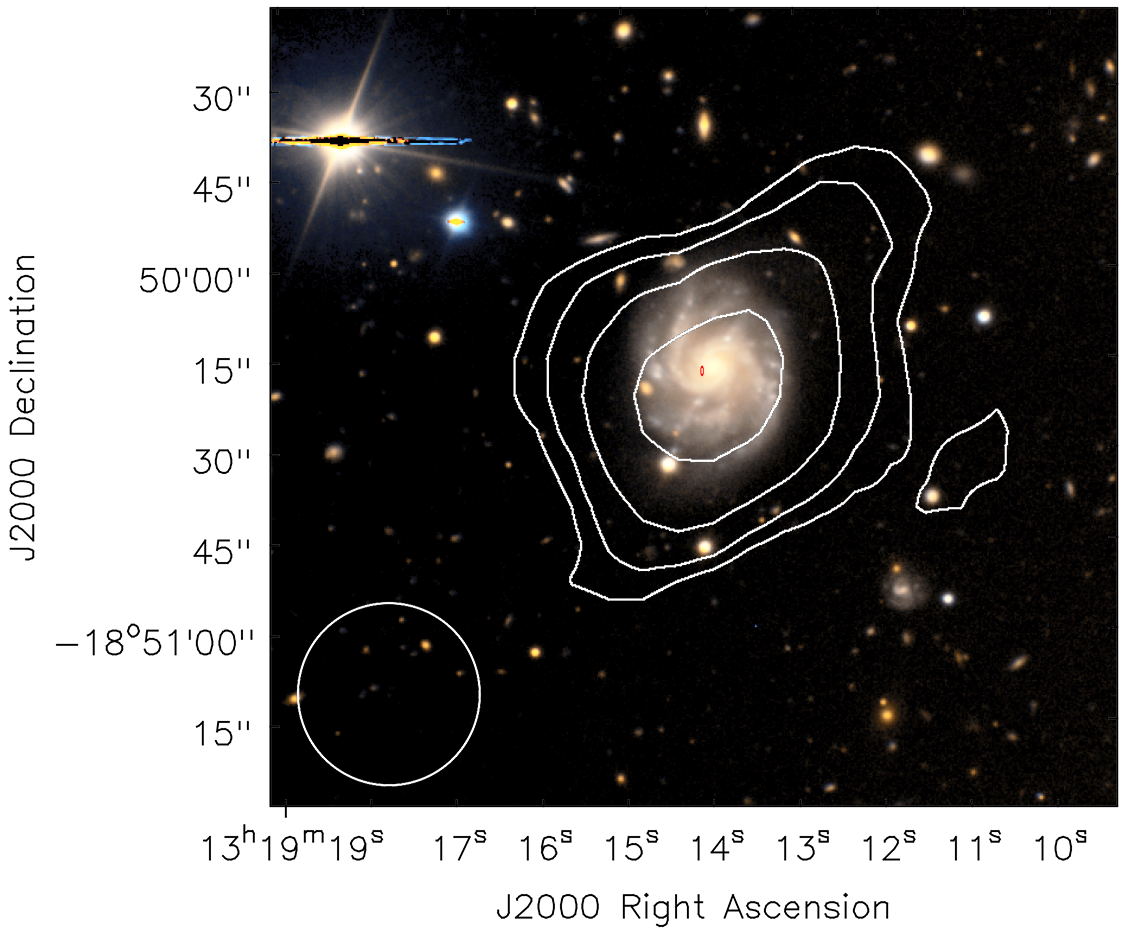}
\caption{\HI\ intensity (moment 0) contour map of W13--18, from the SoFiA source finding output with contour levels shown at 0.6, 1.2, 2.4,  4.8\,$\times$\,10$^{20}$~atoms\,cm$^{-2}$, overlaid on a VLT g-band and I-band image (the RMS of the WALLABY observation corresponds to 0.2\,$\times$\,10$^{20}$~atoms\,cm$^{-2}$). The narrow red ellipse marks the localised position of FRB\,20211127{\myedit I} (Deller et al. in prep). The ASKAP synthesized beam (with spatial resolution = 30$"$) is indicated by the white ellipse in the lower left.}
\label{fig:emissionoverlay}
\end{figure*}

\begin{figure}
\centering
\includegraphics[width=0.9\linewidth]{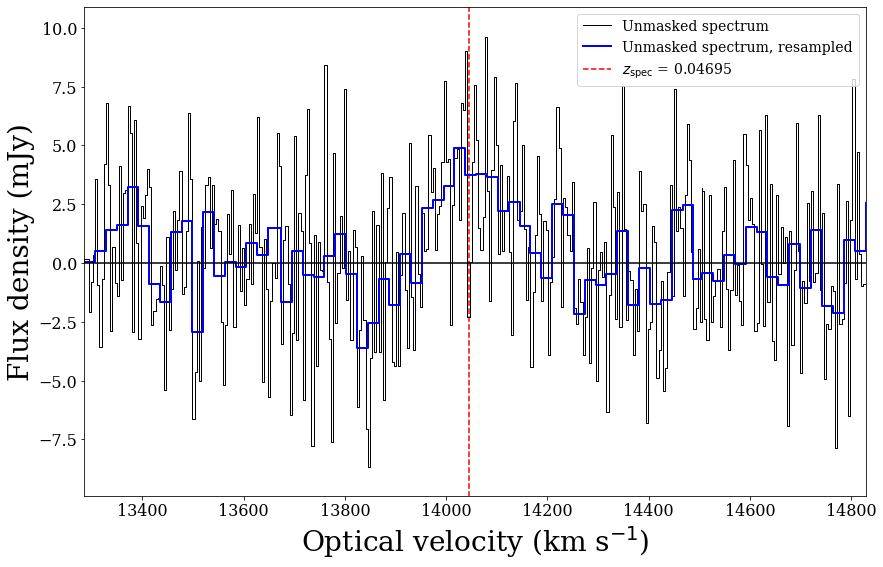}
\caption{\HI\ global spectrum for W13--18 in the heliocentric reference frame. {\myedit The raw \HI\ spectrum is given in black, and a resampled spectrum by a factor of 5 in blue.}
The optical spectroscopic redshift is indicated in red.}
\label{fig:hispectrum}
\end{figure}

\begin{figure}
\centering
\includegraphics[width=0.99\textwidth]{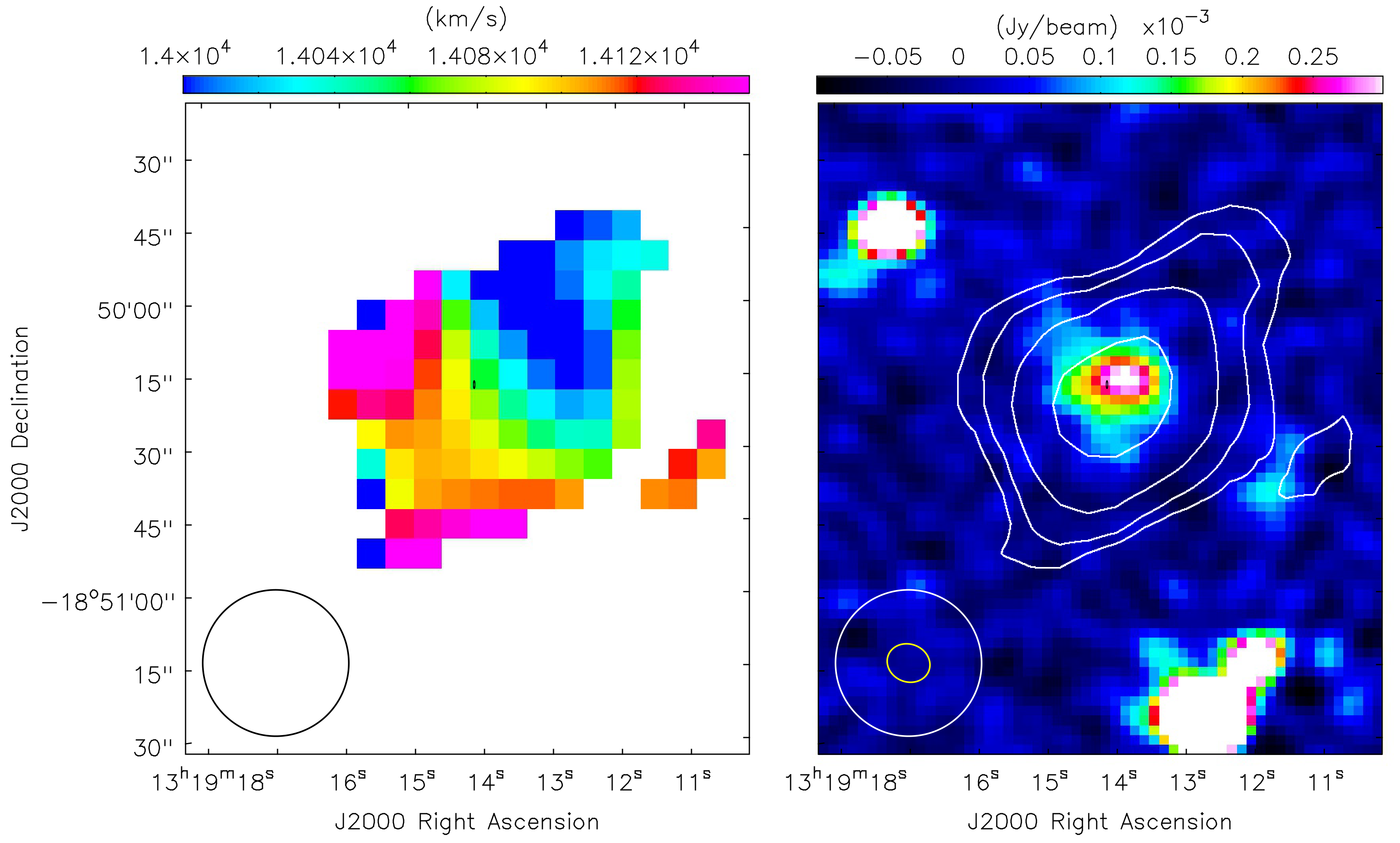}\caption{Left panel: \HI\ velocity (moment 1) map that has been masked using the lowest contour level (0.6\,$\times$\,10$^{20}$~atoms\,cm$^{-2}$) of the moment 0 map. Right panel: \HI\ moment 0 contours (same levels as Fig.~\ref{fig:emissionoverlay}) overlaid on the radio continuum. The 9 arcsec continuum beam, which is at higher resolution than the spectral line cube due to the inclusion of all ASKAP baselines during imaging, is indicated by the smaller yellow ellipse in the bottom left of the panel B. Both panels show the FRB localised position.}
\label{fig:MOM1}
\end{figure}

In Fig.~\ref{fig:emissionoverlay} we present the intensity map for W13--18, displayed as contours overlaid on the VLT image of the FRB host galaxy. The false colour image used the VLT/FORS2 $g$-band for blue, VLT/FORS2 $I$-band for red, and an average of the two optical band images for green, which were combined through astropy's ``make\_lupton\_rgb" function. 
The FRB localisation at 68\% confidence is overlaid as a red ellipse. 
Fig.~\ref{fig:hispectrum} shows the \HI\ global spectrum of the host galaxy, from the raw {\myedit spectral cube, including a version at lower spectral resolution by resampling the spectrum by a factor of 5. We also investigated a SoFiA-masked spectrum that was Hanning-smoothed by a factor of 5.} The velocity resolution of the raw spectrum {\myedit of} the host galaxy is 4.28~km\,s$^{-1}$. In Fig.~\ref{fig:MOM1} we display the intensity map contours overlaid on the velocity map (left panel), and the radio continuum map (right panel). We do not detect any other \HI\ sources along the line of sight to the galaxy in the SoFiA search, nor within the vicinity of the FRB signal. The \HI\ mass sensitivity of the WALLABY observation following equation 157 of \cite{Meyer2017}, assuming a S/N of 5, velocity width of 200~km\,s$^{-1}$, RMS of 1.6~mJy per channel per beam, and an unresolved \HI\ galaxy, is \mhi$_{\rm lim}$ = 2.4~$\times$~10$^{9}$~M$_{\odot}$. The \HI\ mass sensitivity is lower {\myedit near the edges of the footprints}.

We summarise the following properties calculated from the \HI\ data and ancillary data in Table~\ref{tab:galdetails}. Using the optical redshift of $z~=~0.04695$ \cite[][corresponding distance of 215~Mpc assuming Planck cosmology and $H_{\rm 0}$ = 67.7 km\,s$^{-1}$\,Mpc$^{-1}$ \citep{Planck2016}]{Jones2009}, we find \mhi\ = 6.5\,$\pm$1.3~$\times$~10$^{9}$~M$_{\odot}$, where we include an assumed 5\% calibration error for WALLABY in addition to the measured noise ($\pm$1.0~$\times$~10$^{9}$~M$_{\odot}$). Using {\myedit $griz$} photometry and spectroscopy from the Southern Astrophysical Research (SOAR) Telescope (PI Gordon, program SOAR2022B-007){\myedit , $YJKs$ photometry from VISTA \citep{VISTA}, and W1-W4 photometry from WISE \citep{WISE}, \citet{Gordon2023} determines} the SED-derived stellar mass of W13--18 is 3.0~$\times$~10$^{9}$~M$_{\odot}$ (log($M_{*}$/M$_{\odot}$)~=~9.48 $\substack{+0.02 \\ -0.01}$~$M_{\odot}$). This value was calculated using the Bayesian stellar population synthesis code \texttt{Prospector} \citep{Johnson2021}. Further details on the {\myedit data collection, reduction,} assumed priors{\myedit ,} and additional stellar population parameters for this galaxy are reported in {\myedit \cite{Gordon2023}}. Our \HI-to-stellar mass ratio is hence $f_{\rm HI}$~=~$M_{\rm HI}$/$M_{*}$~$\sim$~2.17 (log ratio of 0.34), higher than the 1.3 ratio found for the FRB host studied by \cite{Kaur2022}. 

We attempt to estimate the total dynamical mass, $M_{\rm dyn}$, of W13--18, using equation~2 of \cite{Lee-Waddell2016}. However, due to the low spatial resolution and S/N, we are unable to obtain a reliable estimate.
In order to obtain an improved mass estimate, we attempted to kinematically model this galaxy using both \textsc{3DBarolo}  (3D-Based Analysis of Rotating Objects From Line Observations; \citealt{diTeodoro15}) and a modified version of \textsc{WKAPP} \cite[WALLABY Kinematic Analysis Proto-Pipeline;][]{Deg2022}. Unfortunately the low spatial resolution and $S/N$ precludes such sophisticated modelling. As shown in \cite{Deg2022}, kinematic modelling of WALLABY detections even with full $3D$ codes requires ell\_maj~$\ge2$~beams and $\log(S/N)\ge 1.25$, and this detection satisfies neither requirement.

A radio continuum source with a flux density of 1.3~mJy is associated with the stellar disk at $\sim5\sigma$ significance. The flux-weighted centre of the continuum source, determined through ProFound \citep{Hale2019}, is 13:19:14.06, -18:50:16.0. The angular separation between this source and the FRB localisation is within the image resolution of the synthesised beam (9$"$). Using the ASKAP spectral width at the 50\% flux level of the masked spectrum ($W50$) of 150 km\,s$^{-1}$, we estimate the rotation curve amplitude ($V_{\rm max}$) to be $\sim$100 km\,s$^{-1}$ after correcting for relativistic effects and expected turbulence broadening following \cite{Meyer2008} and also correcting for inclination (19.8\,$^{\circ}$ from 2MASS) following \cite{Meurer2006}. We find this $V_{\rm max}$ to be consistent with the stellar mass estimate for the host galaxy \citep{Wong2016}.

We follow the method described in \cite{Grundy2023} applied to WALLABY galaxies to estimate the total global star formation rate (SFR) from the radio continuum. Using the relationship from \cite{Molnar2021} calibrated against the FIR-radio correlation:

\begin{equation}
    {\rm log_{10}(\frac{SFR_{\rm 1.4}}{M_{\odot}\,yr^{-1}}) = (0.823 \pm 0.009)log_{10}(}\frac{L_{\rm 1.4}}{{\rm W\,Hz^{-1}}}) + 17.5 \pm 0.2,
\end{equation}

we find the SFR to be 2.91$\substack{+1.70 \\ -1.07}$\,$\substack{+0.05 \\ -0.05}$~$M_{\odot}$~yr$^{-1}$. While the WALLABY continuum value is not quite at 1.4~GHz (1.3675~GHz), in the Eridanus pre-pilot WALLABY field the flux values for isolated-unresolved sources are consistent with NVSS at 1.4 GHz within the scatter, so we use WALLABY fluxes without correction. As the origin of the central continuum emission could be due to star formation or a radio-quiet AGN, we also estimate the total SFR beyond the host galaxy core (after subtracting the unresolved core emission) to be 1.77$\substack{+1.04 \\ -0.66}$\,$\substack{+0.07 \\ -0.07}$~$M_{\odot}$~yr$^{-1}$. We present both estimates as it is currently unclear whether the FRB relates to a star-forming region within the core or the inner spiral arm of this host galaxy. The first and second set of uncertainties are the systematic and measured uncertainties for each SFR estimate.

The WISE colours of $W1-W2$~=~0.10\,$\pm$\,0.29~mag and $W2-W3$~=~3.47\,$\pm$\,0.28~mag suggest that W13--18 is unlikely to host a highly efficient accreting AGN as it fits among the normal star-forming spiral region of the WISE colour-colour diagram. As in \cite{Grundy2023}, we calculate the integrated WISE W3PAH (polycyclic aromatic hydrocarbon) SFR, following the relationship given in \cite{Cluver2014}:

\begin{equation}
    {\rm log_{10}(\frac{SFR_{\rm W3PAH}}{M_{\odot}\,yr^{-1}}) = 1.13log_{10}(}v_{\rm W3} * \frac{L_{\rm W3PAH}}{L_{\odot}}) - 10.24,
\end{equation}

calibrated against H$\alpha$ observations. The WISE W3PAH is calculated by subtracting 15.8\% of the total WISE W1 flux from the WISE W3 flux (both measured by ProFound) to account for the contribution of the evolved stellar population to the W3 flux, as in \cite{Jarrett2011, Cluver2017} and references {\myedit therein}. We find the WISE SFR$_{\rm W3PAH}$~=~2.57\,$\pm$\,0.28 $M_{\odot}$\,yr$^{-1}$, which agrees well with the radio continuum SFR.

{\myedit The GALEX near-UV (NUV) magnitude is 17.7345~mag. Converting to a NUV luminosity, we then use equation 6 of \cite{Schiminovich2007},

\begin{equation}
    {\rm SFR_{NUV} ({M_{\odot}\,yr^{-1}}) = 10^{-28.165}}L_{\nu} {\rm (erg s^{-1} Hz^{-1})},
\end{equation}

and find a SFR$_{\rm NUV}$ of 1.22\,$\pm$\,0.05 $M_{\odot}$\,yr$^{-1}$.}

The radio continuum SFR and $M_{\rm HI}$ estimates implies that the star formation efficiency log(SFE$_{\rm HI}$)~=~-10.2$\substack{+0.36 \\ -0.70}$~yr$^{-1}$, where SFE$_{\rm HI}$ is simply SFR/$M_{\rm HI}$. This SFE$_{\rm HI}$ is consistent with the scatter of SFE$_{\rm HI}$ in the sample of \HI-selected star-forming disk galaxies studied by \cite{Wong2016}{\myedit , who argued} that the constant global SFE$_{\rm HI}$ observed for star forming disk galaxies across 5 orders of stellar mass magnitude can be described by self-regulation {\myedit in} a constant marginally-stable disk.

\begin{figure}
\centering
\includegraphics[width=0.8\textwidth]{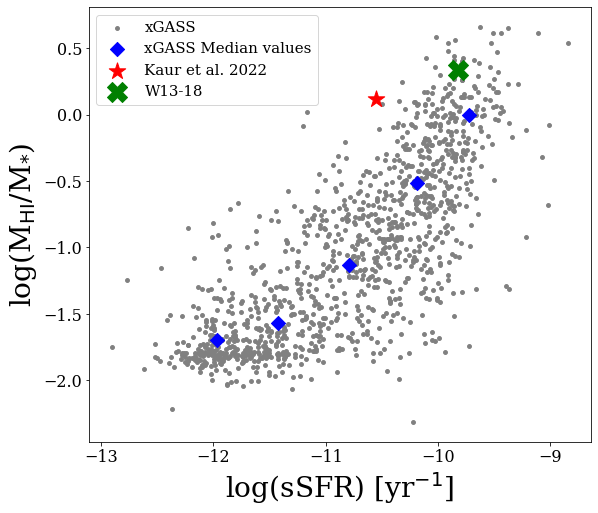}
\caption{The \HI-to-stellar mass ratio, plotted against sSFR, for xGASS galaxies given by the grey points with median in blue diamonds \citep{Catinella2018}, the FRB host studied by \citet{Kaur2022} (SDSS~J0158+6542; red star), and W13--18 (green cross). Both W13--18 and SDSS~J0158+6542 have higher \HI-to-stellar mass ratio, albeit SDSS~J0158+6542 is notably higher than galaxies with comparable sSFR measures to itself.}
\label{fig:xgass}
\end{figure}

From the Prospector SED model, the {\myedit present}-day SFR for W13--18 is 0.45$\substack{+0.59 \\ -0.36}$\,M$_{\odot}$\,yr$^{-1}$ {\myedit \citep{Gordon2023}}. This value describes the SFR of the last 30 Myr, which is sensitive to the youngest stars in the galaxy. {\myedit \cite{Gordon2023} finds that a starburst event may have occurred $\sim$100~Myr ago in W13--18.} 
The optically-derived specific SFR (sSFR) is $\sim$1.5$\times$~10$^{-10}$~yr$^{-1}$ (log value of --9.82). Considering both table~1 and fig. 6 of \cite{Catinella2018}'s analysis of xGASS (the extended GALEX Arecibo SDSS Survey), as done in fig.~2 of \cite{Kaur2022}, W13--18 is gas-rich for galaxies in the mass bin of log$M_{*}$ = 9.44. The xGASS weighted average of logarithm gas fraction is --0.459$\pm$0.067, which is lower than the value we find for W13--18 of 0.34 (at the top of the scatter for xGASS). W13--18 also lies above the log($M_{\rm HI}$/$M_{*}$)-log(sSFR/yr$^{-1}$) relation as the log(sSFR) bin of --9.72 has a corresponding log mass ratio of --0.063$\pm$0.041. The galaxy studied in \cite{Kaur2022} was further above the envelope of the xGASS distribution compared to where W13--18 would lie, at a similar sSFR (Fig~\ref{fig:xgass}). The gas richness of both galaxies may indicate that both FRB hosts recently acquired a large amount of \HI. {\myedit Relative to xGASS, W13--18 lies on the star forming main sequence when considering the optical SFR for the last 30 Myr, or slightly above it when using the total WISE SFR (without subtracting core emission).}

\subsection{Investigating the early trend of strong asymmetry and disturbed HI}\label{sec:comparison}

As discussed in Section~\ref{sec:intro}, other FRB host galaxies found to contain \HI\ gas have shown an early trend: strongly asymmetric \HI\ global spectra and/or disturbed intensity maps indicating these host galaxies are undergoing, or have recently undergone, merger/interaction events. It has been hypothesised that certain merger activity could result in a burst of star formation that in turn could create an FRB progenitor. In these scenarios, the FRB progenitor would be a massive star \citep{Kaur2022} originating from a fast FRB channel model \citep{Michalowski2021}.

For the host galaxy of FRB\,20211127{\myedit I, we are limited by S/N and spatial resolution to confidently state whether this remains the case.} 
The \HI\ intensity map for W13--18 (Fig.~\ref{fig:emissionoverlay}) is fairly featureless and far less disturbed than {\myedit the \HI\ map of SDSS\,J0158+6542} presented in \cite{Kaur2022}, albeit the available spatial resolution is low. The only unusual detail is a \emph{possible} feature to the south west traced by the lower contour of 0.6\,$\times$\,10$^{20}$~atoms\,cm$^{-2}$, with three possibilities. One is that this \HI\ feature is associated with a nearby dwarf galaxy; however, there is no associated source in the optical data, {\myedit and only nearby faint radio continuum emission}. 

The second possibility is the feature is part of a tidal tail connected to the main emission but the spatial resolution is not sufficient to confirm this notion. Furthermore, this feature is on the opposite side of the galaxy from the FRB localisation and would likely not contribute to any enhanced star formation at the FRB source. The third and most likely possibility is that this feature is noise, especially considering its level of significance.

The only indication of any galaxy interaction is a very faint stellar overdensity extending to north-west of the galaxy in the VLT optical image, which appears to correspond with the extension in the \HI\ intensity map. 
The overall position angle and inclination of the \HI\ in W13--18 does not appear to be aligned with the stellar spiral structure; however, higher spatial resolution \HI\ observations at greater sensitivity are required for more detailed analysis.


The \HI\ global spectrum {\myedit has a higher amount of flux on the left hand (approaching) side. 
In order to best compare the level of its asymmetry with FRB hosts studied in} \cite{Michalowski2021}, we adopt the same asymmetry measurement definitions used in that study and the values from \cite{Reynolds2020} , for both the masked and unmasked resampled spectra (bottom panel of Fig.~\ref{fig:hispectrum}). The ratio of the integrated flux in the left and right halves of the spectrum is \cite[see also e.g.][]{Richter1994, Haynes1998, Espada2011}:

\begin{equation}\label{Eq:Aflux}
A_{\rm flux}=|F_{l}/F_{h}|~,
\end{equation}
where 
\begin{equation}
    F_{l}=\int_{v_{l}}^{v_{\rm{sys}}} F(v) dv~,
\end{equation}
and 
\begin{equation}
    F_{h}=\int_{v_{\rm{sys}}}^{v_{r}} F(v) dv~,
\end{equation} 

\noindent where $\nu_{\rm sys}$ is the systemic velocity defined as the mid-point of the spectrum at the 20\% flux level (i.e. $W_{\rm 20}$ line width), and $\nu_{l}$ and $\nu_{r}$ are the left- and right-hand optical velocities where the flux density drops to 20\% of the peak flux density. We also adopt equation 8 of \cite{Reynolds2020}, the residual of the integrated spectrum that was found to have the most significant trend with local galaxy density. Simultaneously, \cite{Deg2020} found the residual of the integrated spectrum to be the best indicator for visually classified asymmetries, as employed in \cite{Glowacki2022}. The integrated spectrum residual is defined as:

\begin{equation}\label{Eq:Aspec}
A_{\rm spec}=\frac{\sum_{i=1}^{}|S(i)-S_{\rm flip}(i)|}{\sum_{i=1}^{}|S(i)|}~,
\end{equation}

\noindent where $S(i)$ and $S_{\rm flip}(i)$ are the fluxes in channel $i$ of the original and flipped spectrum, with the flip axis being the flux-weighted mean systemic velocity. We estimate the error in $A_{\rm flux}$ and $A_{\rm spec}$ in the same way as in \cite{Michalowski2021}: we construct a Gaussian distribution using the standard deviation of line-free channels as the width, and perturbed each channel in the spectrum 1000 times by a value drawn from this Gaussian distribution. The standard deviation of the asymmetry measures of these synthetic spectra was taken as the uncertainty.

\begin{figure}
\centering
\includegraphics[width=0.99\textwidth]{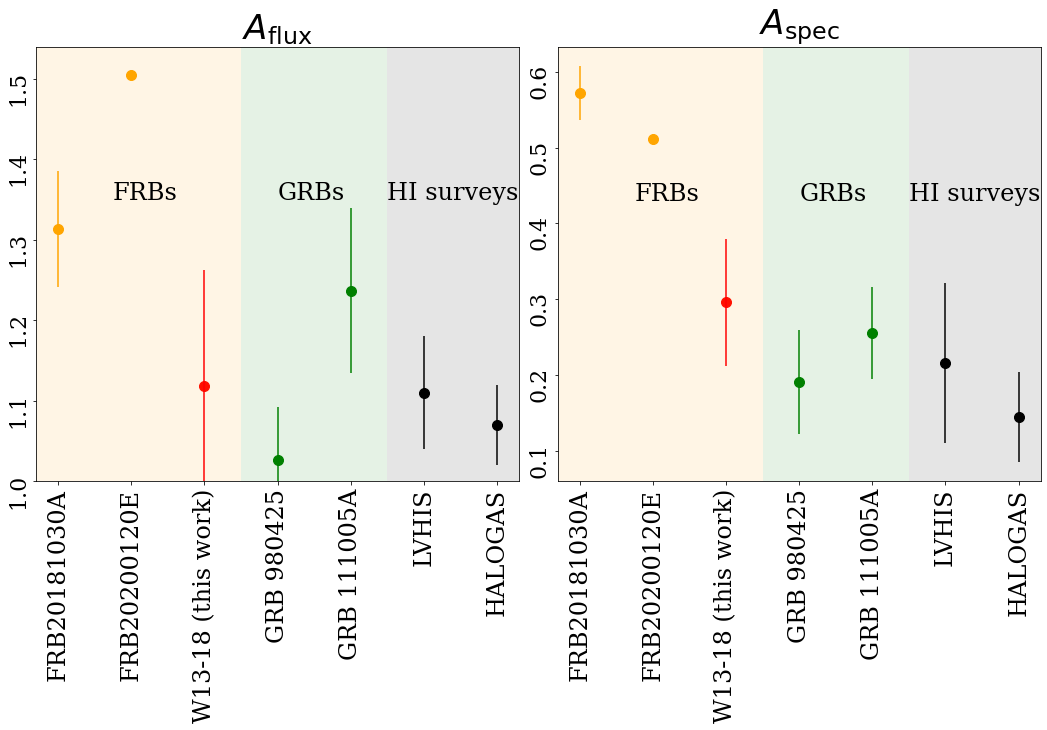}\caption{Comparison of $A_{\rm flux}$ (left panel) and $A_{\rm spec}$ (right panel) for W13--18 with the two FRBs presented in \citet{Michalowski2021} (red and yellow respectively). We include the two GRB hosts (green) and the asymmetry values for the the LVHIS and HALOGAS \HI\ surveys from \citet{Reynolds2020} (black) as in fig.~3 of \citet{Michalowski2021}.}
\label{fig:asymcomp}
\end{figure}

\cite{Michalowski2021} found $A_{\rm flux}$ values of 1.314~$\pm$~0.072 and 1.505~$\pm$~0.002, and $A_{\rm spec}$ values of 0.5719~$\pm$~0.0351 and 0.5108~$\pm$~0.0005 (see table 1 of their study), significantly greater than the values found in \cite{Reynolds2020} for the Local Volume \HI\ Survey \cite[LVHIS;][]{Koribalski2018} and the Hydrogen Accretion in Local Galaxies Survey \cite[HALOGAS;][]{Heald2011} (values of $A_{\rm flux}$~=~1.110~$\pm$~0.070 and 1.070~$\pm$~0.050; and $A_{\rm spec}$ = 0.2160 $\pm$ 0.1060, 0.1450 $\pm$ 0.059 respectively). The unmasked, resampled spectrum of W13--18 has asymmetry values of $A_{\rm flux}$~=~1.12~$\pm$~0.13, and $A_{\rm spec}$~=~0.296~$\pm$~0.091, in agreement with those for the masked resampled spectrum ($A_{\rm flux}$~=~1.14~$\pm$~0.13, $A_{\rm spec}$~=~0.272~$\pm$~0.097). Figure~\ref{fig:asymcomp} compares these values using the unmasked resampled spectrum for W13--18. 

{\myedit In $A_{\rm flux}$ the W13--18 value differs from the FRB hosts examined by \cite{Michalowski2021} by 1.31--2.96$\sigma$, and in $A_{\rm spec}$ 2.36--2.86$\sigma$, not quite sufficient to conclusively state W13-18 does not show the same level of asymmetry as previous hosts. W13-18 is within 0.36$\sigma$ and 1.39$\sigma$ of the mean of the LVHIS and HALOGAS asymmetry values ($A_{\rm flux}$ and $A_{\rm spec}$ respectively). The two FRBs examined by \cite{Michalowski2021}, meanwhile, diverge from those \HI\ surveys by 2.03--8.69$\sigma$ and 2.78--6.22$\sigma$, again in $A_{\rm flux}$ and $A_{\rm spec}$ respectively.}

In addition to the \HI\ global spectrum asymmetries, it is possible to calculate the 2D asymmetry, $A_{2D}$, using the technique of \cite{Conselice2000} and \cite{Conselice2003}, which has been applied in previous studies (e.g.~\cite{Holwerda2011c}) to determine the galaxy merger fraction. The 2D equation is the same as Eqn.~\ref{Eq:Aspec} for $A_{\rm{spec}}$, except the flipping is done in two dimensions and the rotation point is the center of the galaxy. Using a center point of $(22.3, 17.1)$ pixels on the moment 0 map yields a value of $A_{2D}= 0.23$, which is significantly lower than $A_{\rm{spec}}$, but not surprising given the low resolution of the observations \cite[see fig. 12 of][]{Giese2016}.

It is important to consider the effects of noise on both $A_{\rm{spec}}$ and $A_{2D}$. These are channel-by-channel and pixel-by-pixel measurements so noise fluctuations will increase the asymmetry measurement. Noise can also contribute to the lopsidedness measured but since it is an integrated quantity, the effect of noise is decreased (see fig. 3 {\myedit and section 3.2 of \cite{Deg2020}, where a Monte Carlo simulation was performed to explore the effects of S/N on measured asymmetry statistics}).  Thus, the `true' level of asymmetry that can be attributed to the galaxy itself rather than from the noise is lower than our values of $A_{\rm spec}$~=~0.296 and $A_{2D} = 0.23$. 

The mean velocity field for W13--18 (left panel of Fig.~\ref{fig:MOM1}) does not appear to be symmetric, which could be attributed to a warp in the outer disk of W13--18. The outer edges of the velocity field does not have a smooth gradient, particularly on the higher velocity end. The beam size is larger than those high-velocity features and taking into account our limited S/N, we are hence unable to draw any firm conclusions from the velocity field alone.

\cite{Wong2016} argued that self-regulated or secular star formation within disk galaxies across 5 orders of magnitude in stellar mass can be governed by a constant marginally stable disk model. For a galaxy with $V_{\rm max}$ of $\sim$100~km\,s$^{-1}$, this model predicts a SFR of 1.3~$M_{\odot}$\,yr$^{-1}$. {\myedit This is a factor of $\sim$2 lower than the MIR and radio continuum SFR (albeit in agreement within 1$\sigma$ of SFR$_{\rm radio}$ following subtraction of the unresolved core emission), in agreement with the NUV SFR, and higher than the optical SFR in the past 30 years by almost a factor of 3.} 
The nearest WALLABY source to W13--18 is 16.5~Mpc away (down to a $M_{\rm HI}$ sensitivity of 2.4~$\times$~10$^{9}$~M$_{\odot}$), {\myedit suggesting that} this host galaxy is relatively isolated in {\myedit \HI}.

What does it mean for the FRB progenitor? For the previous FRB host galaxies detected in \HI, the characteristics of their \HI\ spectra and intensity maps were attributed to merger/interaction events. However, noting the limitations of the WALLABY data, {\myedit our \HI\ data products for W13--18} do not {\myedit allow us to decide for or against} the hypothesis of a galaxy interaction or merger occurring, which would lead to a recent increase in star formation and the creation of an FRB progenitor (in the form of a massive star). We note that \HI\ asymmetry does not always correspond to a galaxy merger or interaction{\myedit ;} internal processes such as AGN or starburst feedback can also be a potential cause \citep{Sancisi2008,Fraternali2017}. Minor mergers are also not always evident in \HI\ spectra or intensity maps. 


We stress that {\myedit the inability to currently determine a high or low \HI\ asymmetry does} not necessarily argue against
a fast FRB channel model for this FRB progenitor. For instance, the FRB localisation is \emph{not inconsistent} with a spiral arm, which is generally a region of higher active star formation that could possibly be the source of a fast FRB channel progenitor. Furthermore, the high \HI-to-stellar mass ratio observed for the host galaxy, the slight stellar overdensity to the north-west, and the warped velocity field could suggest a minor merger event that could trigger a fast FRB channel for the progenitor. This scenario supports the hypothesis in \cite{Michalowski2021} where the delay between the birth of the progenitor (e.g. a massive star) and the FRB emission is 10--100 Myr, rather than billions of years. However, if such a galaxy interaction event had occurred, it is not {\myedit strongly} reflected in the other \HI\ properties of W13--18, especially compared to FRB host galaxies presented in \cite{Michalowski2021}, \cite{Kaur2022} or Lee-Waddell et al. (submitted). Ultimately, a larger sample of localised FRBs with corresponding host galaxy \HI\ information will be required in order to definitively ascertain the extent to which FRBs are associated with merger-driven star formation.

\section{Conclusions}\label{sec:conclusions}

We present the commensal detection of \HI\ alongside the localisation for FRB\,20211127{\myedit I} to W13--18 through a collaboration between the CRAFT and WALLABY surveys. W13--18 has an \HI\ mass of 6.5\,$\pm$1.3~$\times$~10$^{9}$~M$_{\odot}$. 
Despite the high \HI-to-stellar mass ratio of 2.17, and a possible warp in the \HI\ velocity field, we see no strong disturbances in the \HI\ intensity map, with the only notable feature {\myedit possibly} attributed to noise. {\myedit The} \HI\ intensity map would benefit from an improved S/N and spatial resolution, {\myedit but we nonetheless report the 1D and 2D asymmetry values, which are 1.31--2.96$\sigma$ lower than the FRBs studied in \cite{Michalowski2021} in $A_{\rm flux}$, and 2.36--2.86$\sigma$ lower in $A_{\rm spec}$. 
Higher S/N and spatial resolution \HI\ observations are required to better investigate this FRB host galaxy.}


It is important to remember that W13--18 is only the fifth (and most distant) case of \HI\ emission associated with an external galaxy hosting an FRB. We are still working with low number statistics and are currently treating each detection individually. With a larger sample size, it could be possible that {\myedit we find convincing examples of non-asymmetric \HI\, or find that disturbed \HI\ profiles and intensity maps are the norm for FRB hosts}. As localisations of FRB emission by CRAFT rapidly increase (to $\sim$ 1 per two days) and as \HI\ emission surveys such as WALLABY progress, the sample of \HI\ galaxies that host FRBs should also increase to the point required for a meaningful statistical study (from extragalactic \HI\ surveys with ASKAP, we predict $\sim$5 new \HI\ detections a year as commensally discovered CRAFT FRB host galaxies, not including follow-up studies with other telescopes). Such \HI\ properties for future FRBs can also be used to probe the recent history of the host galaxy, where the star formation history is not available at the time of the FRB detection and localisation.

\section{Acknowledgements}

We thank the referee for useful feedback that has improved this paper. We thank Barbara Catinella, Ronald Ekers, Pascal Elahi, Karl Glazebrook, Kelley Hess, Clancy James, Dane Kleiner, Ángel R. López-Sánchez, Elizabeth Mahony, Gerhardt Meurer, Tom Oosterloo, Javi Román, Lister Staveley-Smith and Lourdes Verdes-Montenegro for useful discussions and feedback on the paper. MG is supported by the Australian Government through the Australian Research Council's Discovery Projects funding scheme (DP210102103). LM acknowledges the receipt of an MQ-RES scholarship from Macquarie University. RMS acknowledges support from the Australian Research Council Future Fellowship FT190100155. SB is supported by a Dutch Research Council (NWO) Veni Fellowship (VI.Veni.212.058). AB acknowledges support from the Centre National d'Etudes Spatiales (CNES), France. JXP as a member of the Fast and Fortunate for FRB Follow-up team acknowledges support from NSF grants AST-1911140 and AST-1910471.

This scientific work uses data obtained from Inyarrimanha Ilgari Bundara, the CSIRO Murchison Radio-astronomy Observatory. We acknowledge the Wajarri Yamaji People as the Traditional Owners and native title holders of the Observatory site. CSIRO’s ASKAP radio telescope is part of the Australia Telescope National Facility (https://ror.org/05qajvd42). Operation of ASKAP is funded by the Australian Government with support from the National Collaborative Research Infrastructure Strategy. ASKAP uses the resources of the Pawsey Supercomputing Research Centre. Establishment of ASKAP, Inyarrimanha Ilgari Bundara, the CSIRO Murchison Radio-astronomy Observatory and the Pawsey Supercomputing Research Centre are initiatives of the Australian Government, with support from the Government of Western Australia and the Science and Industry Endowment Fund.
We also thank the MRO site staff. This research has made use of the NASA/IPAC Extragalactic Database (NED) which is operated by the Jet Propulsion Laboratory, California Institute of Technology, under contract with the National Aeronautics and Space Administration. Parts of this research were supported by the Australian Research Council Centre of Excellence for All Sky Astrophysics in 3 Dimensions (ASTRO 3D), through project number CE170100013. This research is based on observations collected at the European Southern Observatory under ESO programme 0105.A-0687(C) (PI: J.-P. Macquart). We acknowledge support from the Australian SKA Regional Centre (AusSRC).

The stellar mass and associated SFR measures are based on observations obtained at the Southern Astrophysical Research (SOAR) telescope, which is a joint project of the Ministério da Ciência, Tecnologia e Inovações do Brasil (MCTI/LNA), the US National Science Foundation’s NOIRLab, the University of North Carolina at Chapel Hill (UNC), and Michigan State University (MSU).




\footnotesize{
  \bibliographystyle{mnras}
  \bibliography{bibliography}
}


\end{document}